\documentclass[aps,prd,12pt,nofootinbib,
preprintnumbers
,floatfix]{revtex4}
\usepackage{amssymb,hyperref,amsmath}
\usepackage[dvips]{color}
\usepackage[dvips]{graphicx}
\usepackage{epsfig}

\unitlength=1.2mm \preprint{JLAB-THY-07-686}
\begin{document}
\newcommand{\tr}{\mbox{tr}\,}
\newcommand{\Dslash}{{\mathchoice
    {\Dslsh \displaystyle}%
    {\Dslsh \textstyle}%
    {\Dslsh \scriptstyle}%
    {\Dslsh \scriptscriptstyle}}}
\newcommand{\Dslsh}[1]{\ooalign{\(\hfill#1/\hfill\)\crcr\(#1D\)}}
\newcommand{\leftvec}[1]{\vect \leftarrow #1 \,}
\newcommand{\rightvec}[1]{\vect \rightarrow #1 \:}
\renewcommand{\vec}[1]{\vect \rightarrow #1 \:}
\newcommand{\vect}[3]{{\mathchoice
    {\vecto \displaystyle \scriptstyle #1 #2 #3}%
    {\vecto \textstyle \scriptstyle #1 #2 #3}%
    {\vecto \scriptstyle \scriptscriptstyle #1 #2 #3}%
    {\vecto \scriptscriptstyle \scriptscriptstyle #1 #2 #3}}}
\newcommand{\vecto}[5]{\!\stackrel{{}_{{}_{#5#2#3}}}{#1#4}\!}
\newcommand{\vdot}{\!\cdot\!}

\bibliographystyle{apsrev}

\title{$V_{us}$ Calculation from Lattice QCD}

\author{Huey-Wen Lin}
\email{hwlin@jlab.org} \affiliation{Thomas Jefferson National
Accelerator Facility, Newport News, VA 23606}
\date{July, 2007}
\begin{abstract}
I review recent progress in calculating $|V_{us}|$ from lattice QCD kaon and hyperon systems. A preliminary result from the first dynamical calculation in the hyperon channel is included.
\end{abstract}

\maketitle
\section{Introduction}

The Standard Model has successfully described strong interactions using a quantum field theory where universe contains quarks of three generations, which interact via gauge bosons, the gluons. Although electromagnetism and the strong interaction are not affected by quark flavor, the weak interaction may change the flavor of the quarks. This overlap between the various generations is described by the Cabibbo-Kobayashi-Maskawa (CKM) matrix.

In 1963, Cabibbo first introduced a $2\times2$ quark mixing matrix to
explain semileptonic decay in baryons; Kobayashi and Maskawa later
extended the matrix to include the then-undiscovered bottom quark sector.
This becomes the CKM matrix that we are familiar with today:
%
\begin{eqnarray}
V_{\rm CKM} =\left(
\begin{array}{ccc}
V_{ud}    & V_{us}  & V_{ub} \\
V_{cd}    & V_{cs}  & V_{cb} \\
V_{td}    & V_{ts}  & V_{tb} \\
\end{array}
\right).
\end{eqnarray}

The Standard Model requires this matrix to be unitary. This gives six
unitarity constraints, each of which may be graphically depicted as a
unitarity triangle; one such constraint is
\begin{eqnarray}
|V_{ud}|^2 + |V_{us}|^2 + |V_{ub}|^2 = 1.
\end{eqnarray}
From the latest PDG 2006\cite{Yao:2006px}, $V_{ud}$ in this unitarity equation is both the dominant term and also very well determined from neutron beta decay, 0.97377(27); $V_{ub}$ is very small, $4.21(30)\times 10^{-3}$. This leaves the middle matrix element $V_{us}$ as the weak link in deciding whether or not this unitarity equation holds. In latest PDG edition, $V_{us}=0.2257(21)$ is determined to only 0.1\%. In this proceeding, I will describe how this number can be obtained using calculations from lattice QCD. Note that although currently $V_{us}$ seems to indicate that this unitarity equation holds within error bars, the number has been shifting around a lot during the past few years. In 2003, it was 0.2195(23), which is more than 10 standard deviations away from its current best value.

In quantum chromodynamics (QCD), physical observables are calculated from the path integral. For calculations where the coupling is weak, one can perform the integral by hand. However, for long distances the perturbative QCD series no longer converges. Thus, to calculate from first principles, one needs help from lattice gauge theory. Lattice QCD discretizes space-time such that the path integral over field strengths (especially at strong coupling) can be calculated numerically. Since the real world is continuous and infinitely large, at the end of the day we will have to take the lattice spacing $a \rightarrow 0$ and the volume $V \rightarrow \infty$ limits to connect to the physical world.
However, to simulate at the real pion mass (while at the same time
keeping the lattice box big enough to avoid finite-volume effects)
would require much faster supercomputers that have not yet been born.
Thus, we normally calculate with a few unrealistically large values
of the pion mass and then use chiral extrapolation to get back
to the physical pion mass.

Here, we quickly mention a few choices of fermion action that have been
commonly used in lattice QCD calculations. Each has its own pros and cons. They differ primarily by how they maintain symmetry, their calculation cost and their discretization error.
(Improved) Staggered fermions (asqtad)\cite{Kogut:1975ag,Orginos:1998ue,Orginos:1999cr} are relatively cheap for dynamical fermions, but they introduce mixing among parities and flavors or ``tastes'', which make baryonic operators a nightmare to deal with.
The $O(a)$-improved Wilson (clover) fermion action\cite{Sheikholeslami:1985ij} is moderate in cost and free of the disadvantages of the staggered actions. However, chiral symmetry is badly broken at non-zero lattice spacing which causes operator mixing issues.
Chiral fermions (e.g. domain-wall (DWF) or overlap)\cite{Kaplan:1992bt,Kaplan:1992sg,Shamir:1993zy,Furman:1994ky} are free of all the above problems. They are automatically $O(a)$ improved, suitable for spin physics and weak matrix elements. These benefits come at great computational cost.
Last but not least are mixed actions, where one chooses difference valence and sea quark discretizations. Later in this work, we present work done using staggered sea quarks (cheap) with domain-wall valence quarks (chiral); we match the sea Goldstone pion mass to the DWF pion.

\begin{figure}
\includegraphics[width=0.4\textwidth]{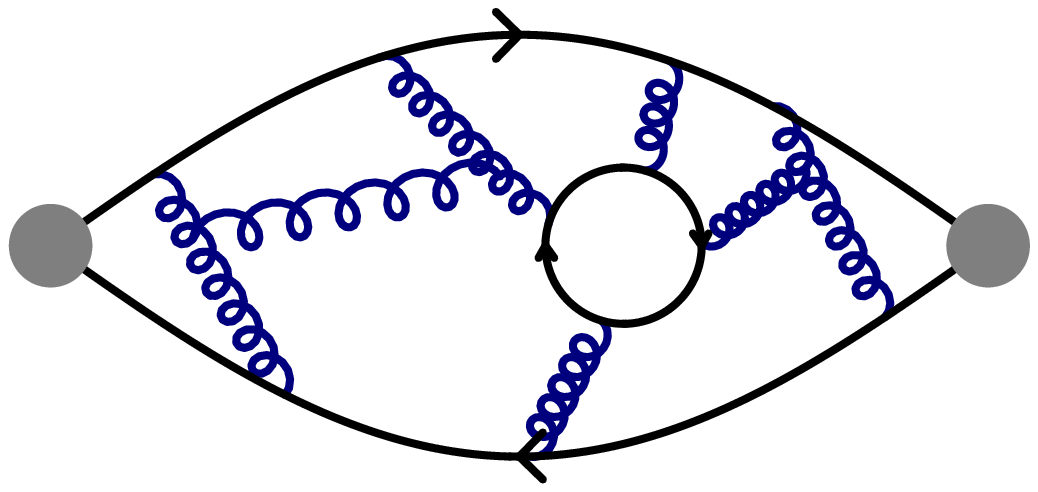}
\includegraphics[width=0.4\textwidth]{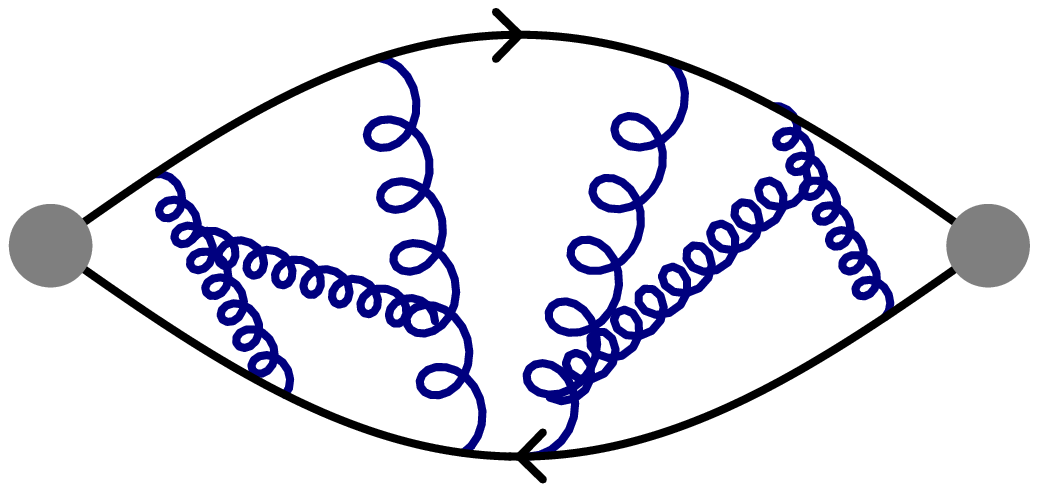}
\caption{A depiction of the difference between full QCD and the quenched approximation in the example of a two-point Green function.}\label{fig:2pt}
\end{figure}

Before we discuss the details of the lattice calculation, it is important to mention the ``quenched'' approximation or 0-flavor calculations. There are a couple of calculations mentioned later which use this approximation. In the path integral, the correct way to carry out the fermionic part of integral is to integrate out the quark/antiquark fields first. This leaves the remaining integral as a function only of the gauge links and introduces a fermionic determinant. The ``quenched'' approximation fixes the fermionic determinant as a constant. This means that when we calculate, say, a two-point Green function, as depicted in Figure~\ref{fig:2pt}, such as a meson correlator, internal fermion loops have been omitted.

The quenched approximation was a product of the old days when computers were slow and algorithms were not yet sped up to today's standards. Quenching very greatly reduces the cost of a lattice calculation by eliminating the fermionic determinant. Of course, modern calculations have moved on to focus on {\it un}quenched calculations. This is a lot better, since when one ignores the fermion loops, it is very difficult to estimate how much this will affect the final numbers; the size of the effect tends to vary greatly between different physical quantities. However, since calculations using the quenched approximation can be done very fast, one can test new methodologies and ideas using this approximation before performing unquenched (dynamical) calculations. The first lattice calculations in $K_{l3}$ and hyperon semileptonic decay were demonstrated in this approximation.

In the following, I will review the latest $V_{us}$ calculations. There has been a series of works devoted to $K_{l3}$ decays but only quenched calculations in the hyperon channel so far. In the second half of this work, I will show the first lattice dynamical calculation of hyperon decays using mixed action. In the final part, I will summarize the current standing of $V_{us}$ from lattice QCD and give some future outlook.

\section{Lattice $V_{us}$ Calculations}\label{sec:lattice Vus}
In this work, we will concentrate on three determinations of
$V_{us}$:
leptonic decay ratios (mainly for completion), $K_{l3}$ decays and hyperon decays. We will compare all of them to the number listed in PDG 2006. So far, the number from $K_{l3}$ decays has the smallest errorbar for $V_{us}$.

\subsection{Leptonic Decays}
If one looks at the decays $K_{\mu2}$ and $\pi_{\mu2}$, their branching ratios can be written in terms of $V_{us}/V_{ud}$ and the ratio of the kaon to pion decay constant:
\begin{equation}
\left( \frac{\left|V_{us}\right|}{\left|V_{ud}\right|} \right)^2 =
  \left[ \left( \frac{f_K}{f_\pi} \right)^2
           \frac{M_K \left(1-m_\mu^2/M_K^2\right)^2}
                {M_\pi \left(1-m_\mu^2/M_\pi^2\right)^2}
        \left( 1+\frac{\alpha}{\pi}\left(C_K-C_\pi\right) \right) \right]^{-1}
  \frac{\Gamma(K \rightarrow \mu \bar{\nu}_\mu)}
       {\Gamma(\pi \rightarrow \mu \bar{\nu}_\mu)},
\end{equation}
where $C_K$ and $C_\pi$ are the radiative-inclusive electroweak corrections, and the rest of the numbers can be obtained from experimental measurements.
W.~J.~Marciano\cite{Marciano:2004uf} used decay constant ratios $f_K/f_\pi=1.210(4)(13)$ from a 2+1f staggered fermion calculation of $f_\pi$ and $f_K$ done in 2004 by MILC collaboration and found $V_{us}=0.2219(25)$. Of course, there have been other full-QCD calculations since 2004. For example, RBC/UKQCD use dynamical chiral fermions (DWF) to obtain the ratio 1.24(2)\cite{Allton:2007hx}. However, none of the other collaborations have come out with a number with competitive errorbar yet. In 2006, MILC updated their own calculation, $1.208(2)(^7_{14})$\cite{Bernard:2006wx}; this yields $V_{us}=0.2223(^{26}_{14})$.

\subsection{$K_{l3}$ Decay}
\begin{figure}
\includegraphics[width=0.5\textwidth]{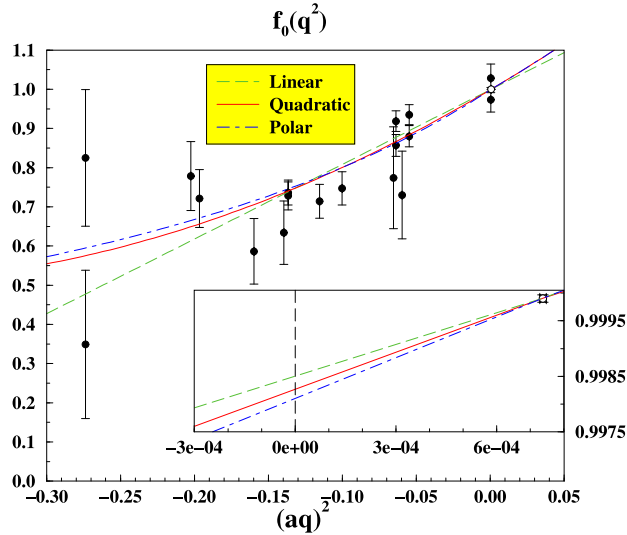}
\caption{Momentum extrapolation from Ref.~\cite{Becirevic:2004bb} }\label{fig:f0_Damir}
\end{figure}

Another way to determine $V_{us}$ is to look at the $K_{l3}$ decay.
When one integrates out the short-distance dependence, one is left with a low-energy non-perturbative matrix element for K to $\pi$, which can be calculated directly in lattice QCD. Using Lorentz invariance, we can decompose the matrix element into two form factors ($f_+$ and $f_-$) with differing momentum dependence:
\begin{equation}
\langle \pi(p^\prime)|V_\mu|K(p)\rangle = (p_\mu+p^\prime_\mu) f_+(q^2)+ (p_\mu-p^\prime_\mu) f_-(q^2).
\end{equation}
The so-called ``double ratio'' technique:
\begin{equation}\label{eq:doubelR}
\frac{\langle \pi|\overline{s}\gamma_0u|K\rangle
    \langle K|\overline{u}\gamma_0s|\pi\rangle}
{\langle K|\overline{s}\gamma_0s|K\rangle
    \langle \pi|\overline{u}\gamma_0u|\pi\rangle }
    = |f_0(q_{\rm max}^2)|^2 \frac{(m_K+m_\pi)^2}{4m_Km_\pi},
\end{equation}
which has been used to look at $B$-to-$D$ decays\cite{Hashimoto:1999yp}, can also be applied to the $K$ to $\pi$ decay. The result is a form factor that only depends on $q^2$, the momentum transfer between the initial and final states, and some kinetic factors. This $f_0$ can be connected to $f_\pm$ through
\begin{equation}
f_0(q^2)=f_+(q^2)+\frac{q^2}{m_K^2-m_\pi^2}f_-(q^2).
\end{equation}
When we extrapolate to $q^2=0$, $f_0 = f_+$. We study different extrapolation forms to estimate the systematic error caused by our extrapolation:
\begin{eqnarray}
f_0(q^2)^{\rm Linear}    &=& f_0(0) (1+\lambda_0q^2) \\
f_0(q^2)^{\rm Quadratic} &=& f_0(0) (1+\lambda_0q^2+c q^4) \\
f_0(q^2)^{\rm Polar}     &=& f_0(0)/(1-\lambda_0q^2),
\end{eqnarray}
which are shown in Figure~\ref{fig:f0_Damir} by Becirevic et al.\cite{Becirevic:2004bb}.

To remove uncertainty due to the momentum extrapolation, we can calculate the matrix element directly at $q^2=0$. The trick to this technique is to include help from ``twisted'' boundary conditions on the fermions as
\begin{equation}
\psi(x+e_jL)=e^{2\pi i \theta_j}\psi(x).
\end{equation}
The discretized momenta on the lattice will reflect the choice of this $\theta_j$ as $p_j=\theta_j\frac{2\pi}{L}+n_j\frac{2\pi}{L}$ with $n_j$ integer. We can select $\theta$ wisely to cancel out the mass difference, so that we can obtain $f_+(0)$ directly. Guadagnoli~et~al.\cite{Guadagnoli:2005be} first demonstrated the advantage in the quenched approximation. Later, UKQCD made an exploratory study on a 2+1-flavor DWF calculation\cite{Boyle:2007wg}. They showed that with 50\% more statistics, one can get number competitive with conventional extrapolation calculations. The advantage of this method over the conventional one is smaller or no systematic error due to $q^2$ extrapolation. Thus, total error on the calculation is reduced.

After we obtain $f_0(0)$, the next step is to extrapolate the pion mass to the physical one. We can get some help from the Ademollo-Gatto (AG) theorem\cite{Ademollo:1964sr,Becirevic:2005py}. We know that the $SU(3)$ symmetry-breaking Hamiltonian is
\begin{equation}
H=\frac{1}{\sqrt{3}}\left(m_s-\frac{m_d+m_u}{2}\right)\overline{q}\lambda^8q.
\end{equation}
The AG theorem tells us that there is no first-order correction due to
$SU(3)$-breaking; thus, the correction starts at second order
\begin{equation}
f_0(0)=f_0(0)^{SU(3)}+O(H^2).
\end{equation}
What would be a good measure for $SU(3)$ breaking? The most natural candidate would be the mass splitting between the kaon and pion. So, we expect the remaining correction should be small; thus, one would expect the ``corrected'' lattice $f_0$ (after subtracting the chiral log),
$f^\prime$, should differ from $f_0^{SU(3)}$ by only a small amount. We construct a ratio
\begin{equation}\label{eq:AG-R}
R(m_K,m_\pi) =\frac{f^{SU(3)}-|f^\prime(0)|}{a^4(m_K^2-m_\pi^2)^2},
\end{equation}
where $f^{SU(3)}$ is the $SU(3)$-limit value; in this case, it is 1. We expect the remaining mass dependence in Eq.~\ref{eq:AG-R} should be relatively small. We then extrapolate the remaining mass dependence to the physical sum of the pion and kaon masses-square
\begin{equation}\label{eq:AGR-exp}
R(m_K,m_\pi) = c_0 +c_1 {a^2(m_K^2+m_\pi^2)}.
\end{equation}
Thus, $V_{us}$ can then be obtained from
\begin{equation}
\Gamma(K_{l3}) = \frac{G_F^2M_K^5}{128\pi^3}|V_{us}|^2
S_{\rm EW}|f_+(0)|^2
C_K^2 I_K^l(\lambda_i)
(1+\delta_{SU(2)}^K+\delta_{\rm EM}^{K}).
\end{equation}
The decay width, $\Gamma(K_{l3})$, is taken from experiment,
while the phase-space integral $I_K^l$,
isospin breaking $\delta_{SU(2)}^K$,
long-distance electromagnetic corrections $\delta_{\rm EM}^{K}$ and
short-distance radiative corrections $S_{\rm EW}$ are taken from perturbative calculations.

\begin{figure}
\includegraphics[width=0.5\textwidth]{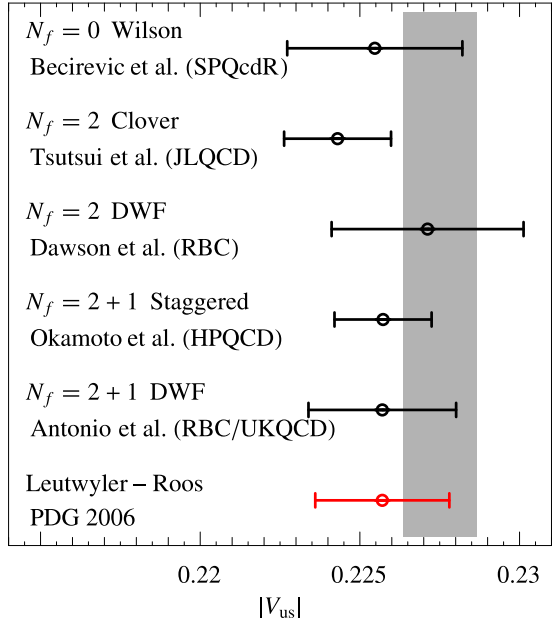}
\caption{Lattice $|V_{us}|$ summary with unified experimental numbers from PDG}\label{fig:VusKl3}
\end{figure}

Table~\ref{tab:Kl3_Vus} summarizes the results from various lattice QCD groups: quenched, partially quenched and full QCD, fermion action variety and the range of the pion mass. Figure~\ref{fig:VusKl3} $f_+$ is taken from individual calculations, combined with the latest PDG 2006 number for $|f_+ V_{us}|=0.2169(9)$. Note the $V_{us}$ number may be different from the ones given in the original papers, due to progress in experimental measurements. The grey band in the graph is the range allowed assuming unitarity holds. All the lattice calculations so far have agreed with the old estimation from Leutwyler-Roos in 1984\cite{Leutwyler:1984je}.
However, not every paper has complete estimations of the systematic errors due to lattice artifacts. The work done by RBC/UKQCD\cite{Antonio:2007mh} is one of the exceptions, and thus I quote their number as representative of the $V_{us}$ from $K_{l3}$ decay channel: 0.2257(14).

\begin{table}
\begin{center}
\begin{tabular}{|c|c|c|c|c|c|}
\hline
Group & $N_{\rm f}$ & $S_{\rm f}$ & $M_\pi$ (GeV) & \# conf & $f_+(0)$ \\
\hline \hline
SPQcdR\cite{Becirevic:2004ya}    & 0   & Wilson    & 0.500--1.000 &  230  & 0.961(09) \\
%
JLQCD\cite{Tsutsui:2005cj}     & 2   & Clover    & 0.440--$0.960^*$& N/A 
& 0.967(06) \\
%
RBC\cite{Dawson:2006qc}      & 2   & DWF       & 0.475--0.700 &   94 & 0.955(12) \\
%
HPQCD\cite{Okamoto:2005zg}     & 2+1 & Staggered & 0.500--0.700 & N/A 
& 0.962(11) \\
%
RBC/UKQCD\cite{Antonio:2007mh} & 2+1 & DWF       & 0.390--0.700 &  150 & 0.961(05) \\
%
\hline
\end{tabular}
\end{center}
\caption{Summary of existing published $f_+$ calculations from $K_{l3}$ decay}\label{tab:Kl3_Vus}
\end{table}

\subsection{Hyperon Decays}
Hyperon decays provide us with an additional independent channel for determining $V_{us}$. We start by looking at the low-energy contribution of the transition matrix elements for hyperon beta decay,
$B_1 \rightarrow B_2 e^- \overline{\nu}$; in low-energy effective theory this can be written as
\begin{eqnarray}\label{eq:ME}
{\cal M} =\frac{G_s}{\sqrt{2}}\overline{u}_{B_2}(O_\alpha^{\rm
V}+O_\alpha^{\rm
A}){u}_{B_1}\overline{u}_{e}\gamma^\alpha(1+\gamma_5)v_{\nu}.
\end{eqnarray}
From Lorentz symmetry, we expect the matrix element composed of any two spin-$1/2$ nucleon states, $B_1$ and $B_2$, to have the general form
\begin{eqnarray}\label{eq:cont_O's}
O_\alpha^{V} &=& f_1(q^2)\gamma^\alpha
+\frac{f_2(q^2)}{M_{B_1}}\sigma_{\alpha\beta}q^\beta
+\frac{f_3(q^2)}{M_{B_1}}q_\alpha \\
O_\alpha^{A} &=& \left(g_1(q^2) \gamma^\alpha
+\frac{g_2(q^2)}{M_{B_1}}\sigma_{\alpha\beta}q^\beta
+\frac{g_3(q^2)}{M_{B_1}}q_\alpha\right)\gamma_5
\end{eqnarray}
with transfer momentum $q=p_{B_2}-p_{B_1}$ and $V,A$ indicating the vector and axial currents respectively.

The vector form factor is connected to $V_{us}$ via
\begin{eqnarray}\label{eq:hyperon_Vus}
\Gamma & = & G_F^2 \, |V_{us}|^2 \, \frac{\Delta m^5}{60 \pi^3} \,
(1+\delta_{\rm rad}) \\*[1.5mm]
       &  &  \!\!\!\!\!\!  \times \left[ \left( 1 - \frac{3}{2} \, \beta \right)
       \left( |f_1|^2 + |g_1|^2 \right)
       + \frac{6}{7} \, \beta^2 \left( |f_1|^2 + 2 |g_1|^2 +\text{Re}(f_1 f_2^*)
       + \frac{2}{3} \, |f_2^2| \right)
       + \delta_{q^2} \right],
\end{eqnarray}
with $\Delta m = m_{B_1} - m_{B_2}$, $\beta = \Delta m / m_{B_1}$, the radiative corrections $\delta_{\rm rad}$, and $\delta_{q^2}(f_1,g_1)$ taking into account the transfer-momentum dependence of $f_1$ and
$g_1$~\cite{Garcia:1985xz}.
Generally, the ratios of $g_1/f_1$ from experiment and $f_2/f_1$ in the $SU(3)$ limit are used to get $V_{us}$ from hyperon decays.

In 2003, Cabibbo~et~al.\cite{Cabibbo:2003cu} used $f_2/f_1$ and $f_1$ in the $SU(3)$ limit, combined with the experimental decay width (or rate) and $g_1/f_1$, to obtain $V_{us}$ from various channels of hyperon decay, as shown in Table~\ref{tab:Cabibbo}. It is not hard to see that if lattice calculations can provide better estimates of $g_1/f_1$, we can improve the precision of $V_{us}$ from hyperon decays and possibly get a better estimation than the $K_{l3}$ channel.

\begin{table}
\begin{center}
\begin{tabular}{c|ccccc}
\hline\hline
    Channel         & $f_1^{SU(3)}$ & $|f_1V_{us}|$ & $\left({g_1}/{f_1}\right)^{SU(3)}$ & $\left({g_1}/{f_1}\right)^{\rm exp}$  \\
\hline
$n \rightarrow p$             & 1            & $n/a$        & $F+D$  & $1.2670(30)$&\\
$\Lambda \rightarrow p$       &$-\sqrt{3/2}$ & $0.2221(33)$ & $F+D/3$& $0.718(15)$&\\
$\Sigma^{-} \rightarrow n$    &$-1$          & $0.2274(49)$ & $F-D$  & $-0.340(17)$&\\
$\Xi^{-} \rightarrow \Lambda$ &$\sqrt{3/2}$  & $0.2367(97)$ & $F-D/3$& $0.25(5)$&\\
$\Xi^{-} \rightarrow \Sigma^0$&$\sqrt{1/2}$  & $n/a$        & $F+D$  & $n/a$&\\
$\Xi^{0} \rightarrow \Sigma^+$& 1            & $0.216(33)$  & $F+D$  & $1.32(22)$&\\
\hline\hline
\end{tabular}
\end{center}
\caption{Summary of a few hyperon numbers}\label{tab:Cabibbo}
\end{table}

So far, there are only two quenched lattice calculations of hyperon beta decay, and they are in different channels, $\Sigma \rightarrow n$ and $\Xi^0 \rightarrow \Sigma^+$. Guadagnoli~et~al.\cite{Guadagnoli:2006gj} extrapolate the matrix element $\Sigma \rightarrow n$ via an AG ratio, similar to the discussion in the $K_{l3}$ decay case, but using a dipole form to extrapolate to the zero-transfer momentum point. All of the pion masses are larger than 700~MeV; their final numbers are
$f_1=-0.988(29)_{\rm stat}$ and $V_{us}= 0.230(5)_{\rm exp}(7)_{\rm lat}$.
Sasaki~et~al.\cite{Sasaki:2006jp} use lighter pion masses 530--650~MeV and DWF to look at the $\Xi^0$ decay channel. They extrapolate the vector form factor $f_1$ via the variable $\delta=(m_{B_2}-m_{B_1})/m_{B_2}$.
The Ademollo-Gatto theorem suggests the leading-order effect should be $\delta^2$, and thus one can fit $f_1(0)$ to the form $c_0+c_1\delta^2$. Their final numbers are $f_1=0.953(24)_{\rm stat}$ and $V_{us}= 0.219(27)_{\rm exp}(5)_{\rm lat}$.
Unfortunately, the experimental determination of the decay rate is lousy; despite $f_1$ in $\Xi$ decay channel being compatible within errors, $V_{us}$ is not well-determined. This may further improve in the future with updates from Fermilab KTeV and CERN NA48 collaborations. One important thing to note is that neither of the calculations has systematic error estimates from quenching effects, which we expect might be significant.

We have taken data looking at both hyperon channels with a dynamical
lattice calculation for the first time. We use a mixed action, meaning that the sea (staggered) and valence (DWF) fermions have different discretization. Our pion masses are relatively lighter than the quenched calculations. We only simulate one strange quark mass, which unfortunately does not reproduce the correct strange-strange Goldstone mesons. We find a box size around 2.6~fm and list a few other important parameters in Table~\ref{tab:conf_Info}. In this work, we report on our preliminary calculation in the $\Sigma^- \rightarrow n$ channel. We use a projection operator $T=(1-\gamma_5\gamma_3)(1+\gamma_4)/2$ in both two- and three-point Green functions and construct a ratio
\begin{eqnarray}\label{eq:Ratio_GP}
R_{j_\mu} &=& \frac{Z_V
      \Gamma^{\Sigma N}_{\mu,GG}(t_i,t,t_f,\vec p_i,\vec p_f\,;\;T)}{\Gamma^{NN}_{GG}(t_i,t_f,\vec p_f\,;\;T)}
 \sqrt{\frac{\Gamma^{\Sigma\Sigma}_{PG}(t,t_f,\vec
p_i\,;\;T)}{\Gamma^{NN}_{PG}(t,t_f,\vec
      p_f\,;\;T)}}\nonumber \\
 &\times&     \sqrt{\frac{\Gamma^{NN}_{GG}(t_i,t,\vec p_f\,;\;T)}{\Gamma^{\Sigma\Sigma}_{GG}(t_i,t,\vec
      p_i\,;\;T)}}
   \sqrt{\frac{\Gamma^{NN}_{PG}(t_i,t_f,\vec
p_f\,;\;T)}{\Gamma^{\Sigma\Sigma}_{PG}(t_i,t_f,\vec
      p_i\,;\;T)}},
\end{eqnarray}
to cancel out kinetic and overlap $Z$ factors. With multiple insertions of the momentum, we can solve for the individual form factors in Eq.~\ref{eq:cont_O's}.

\begin{table}
\begin{center}
\begin{tabular}{c|ccccc}
\hline\hline
Label & $m_\pi$ (MeV) &  $m_K$ (MeV) & $\Sigma^- \rightarrow n$ conf.
\\
\hline
m010  & 358(2)        & 605(2) & 600  \\
m020  & 503(2)        & 653(2) & 420  \\
m030  & 599(1)        & 688(2) & 561  \\
m040  & 689(2)        & 730(2) & 306  \\
\hline\hline
\end{tabular}
\end{center}
\caption{Configuration information}\label{tab:conf_Info}
\end{table}

\begin{figure}
\includegraphics[width=0.7\textwidth]{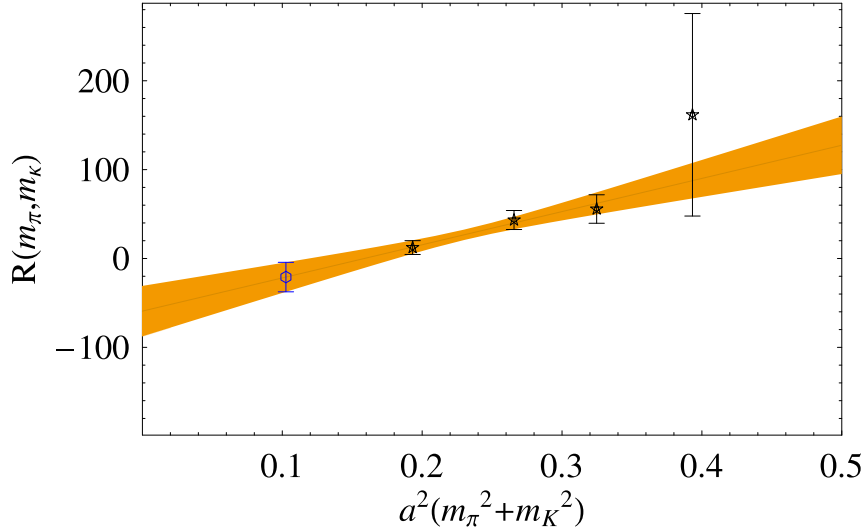}
\caption{AG ratio extrapolation to physical $m_K^2+m_\pi^2$}\label{fig:fmPK}
\end{figure}

We need to extrapolate to zero momentum. We use a dipole form,
as has been used in momentum extrapolation for many baryons' momentum
dependence. For the mass extrapolation, a similar approach to
the $K_{l3}$ case can be applied here with the help of the Ademollo-Gatto
theorem. We first construct a ratio and then extrapolate the mass dependence according to Eq.~\ref{eq:AG-R}, as shown in Figure~\ref{fig:fmPK}.
Since our heaviest pion mass is much closer to the strange Goldstone
meson mass, due to the mass difference in the denominator of the ratio,  the magnitudes of this point, both central value and especially the errorbar, increase a lot. Thus, it provides only a very weak constraint to the fit; this is not an ideal solution for our data.

\begin{figure}
\includegraphics[width=0.7\textwidth]{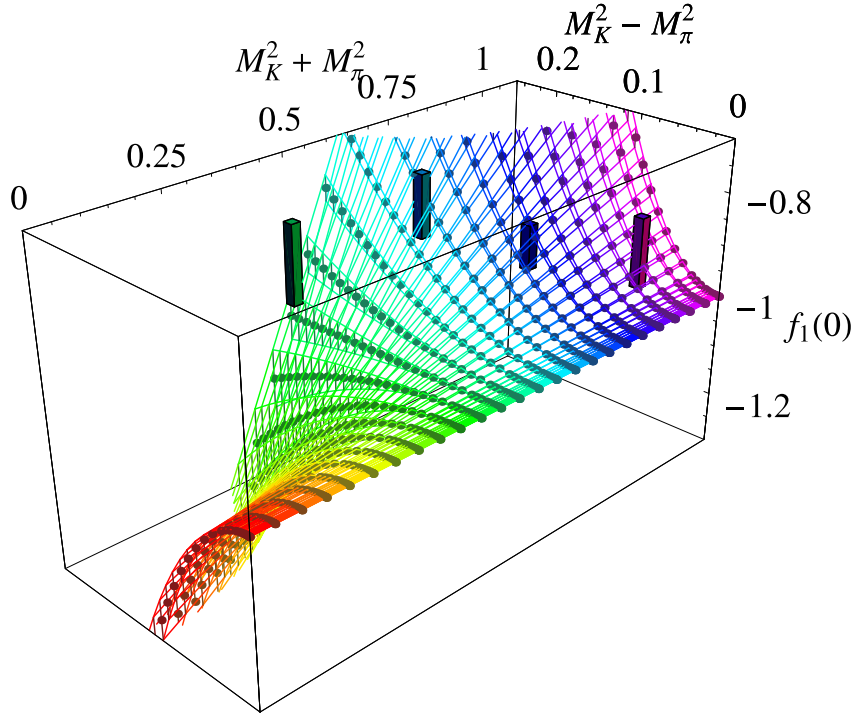}
\caption{Two-dimensional mass extrapolation after dipole extrapolation to zero-transfer point}\label{fig:f2dmkmp}
\end{figure}

Alternatively, we can combine the given two-step process into one, by performing a two-dimensional fit to the sum and difference of kaon and pion masses, as shown in Figure~\ref{fig:f2dmkmp}. But this fit is not
ideal either since the constraints from the data points are not strong: only four points to define a two-dimensional surface.

\begin{figure}
\includegraphics[width=0.7\textwidth]{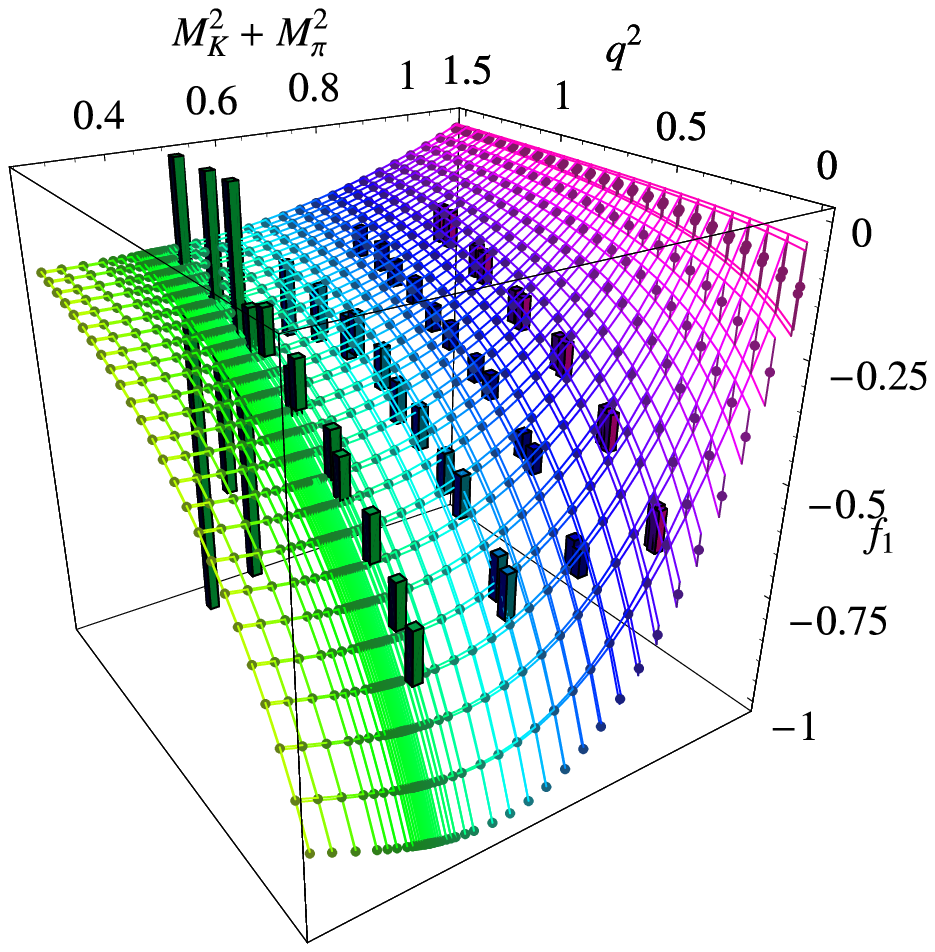}
\caption{Simultaneous extrapolation in $q^2$ and mass }\label{fig:f2dmq}
\end{figure}

A better constrained fit can be composed by combining the momentum and mass dependence into a single simultaneous fit:
\begin{equation}\label{eq:CombinedFit}
f_1(q^2) = \frac{1+\left(M_K^2-M_\pi^2\right)^2
\left(A_1+A_2\left(M_K^2+M_\pi^2\right)\right)}
{\left(1-\frac{q^2}{M_0+M_1\left(M_K^2+M_\pi^2\right)}\right)^2}.
\end{equation}
Figure~\ref{fig:f2dmq} shows the result from simultaneously fitting over all $q^2$ and mass combinations. The $z$-direction indicates $f_1$, while the $x$- and $y$-axes indicate mass and transfer momentum. The surface is the fit using Eq.~\ref{eq:CombinedFit} with color to indicate different masses. The columns are the data and the momentum points from different pion masses line up in bands.
Our preliminary result for $f_1$ is $-0.88(15)$. This leads us to a $V_{us}$ somewhat larger in central value than the other calculations but still agrees with them due to the large errorbar. The statistics will be greatly improved at the lightest pion mass data in the near future.

\section{Conclusion and Outlook}
To summarize, there are various ways that lattice QCD calculations can help to determine $V_{us}$ in the CKM matrix. (Similar approaches can be applied to the rest of the elements with effective lattice fermion actions.) Firstly, we can use the lattice input from the kaon and pion decay constant ratios. Currently, MILC has best determined ones, resulting in $V_{us} = 0.2226(^{26}_{15})$. Secondly, we can use the form factor from $K_{l3}$ decay matrix elements: here we use the number from RBC/UKQCD, 0.2257(14), in which a sound study and proper systematics are included.

Finally, we can use the form factors from hyperon decays. We have started the first full-QCD 2+1-flavor dynamical calculation. Our preliminary results show consistency with previous calculations, but have larger errorbar due to the choice of lighter pion mass. The larger statistical error is partially compensated by the decrease in systematic error due to extrapolating the pion mass to the physical one. To improve the $V_{us}$ value from the hyperon decays, we need to reduce our statistical error on  the vector form factor and improve the accuracy on $g_1/f_1$ to replace the experimental one. Using these strategies, we can make our calculation of $|V_{us}|$ equivalent to or better than the one from the $K_{l3}$ channel.

\section*{Acknowledgements}
HWL thanks collaborator Kostas Orginos for useful discussions on hyperon decays. Computations were performed using the Chroma software suite\cite{Edwards:2004sx} on clusters at Jefferson
Laboratory using time awarded under the SciDAC Initiative.  Authored by Jefferson Science Associates, LLC under U.S. DOE Contract No. DE-AC05-06OR23177. The U.S. Government retains a non-exclusive, paid-up, irrevocable, world-wide license to publish or reproduce this manuscript for U.S. Government purposes.

\bibliography{exp}

\end{document}